\documentclass[preprint,prd,amsmath,amssymb,axodraw,showpacs]{revtex4}

\usepackage{epsfig}
\usepackage{dcolumn}
\usepackage{bm}
\usepackage{amssymb, amsmath, amsthm}
\usepackage{graphicx}
\DeclareMathOperator{\Tr}{Tr}

\begin{document}



\title{Charm fragmentation functions in the Nambu-Jona-Lasinio model}


\author{Dong-Jing Yang}\email{djyang@std.ntnu.edu.tw}
\affiliation{Institute of Physics, Academia Sinica, Taipei 11529, Taiwan, Republic of China}
\author{Hsiang-nan Li}\email{hnli@phys.sinica.edu.tw}
\affiliation{Institute of Physics, Academia Sinica, Taipei 11529, Taiwan, Republic of China}


\date{\today}

\begin{abstract}
We derive the fragmentation function (FF), which describes the probability for a charm quark to emit a $D$ meson with a certain momentum fraction, in the Nambu-Jona-Lasinio (NJL) model. The corresponding elementary FF
is calculated with the quark-meson coupling determined in the NJL model involving charm quarks. The FF in the infinite momentum frame is constructed through the jet process governed by the elementary FF, and then evolved to the charm scale, at which it is defined. To prepare the FF suitable for an analysis of $D$ meson production at CLEO, we further match the above FF to that in the finite momentum frame at one loop in QCD.
It is shown that the charm quark FF including the finite momentum effects leads to theoretical results in agreement with the CLEO data.
\end{abstract}


\maketitle


\section{Introduction}

Parton fragmentation functions (FFs) contain important information on strong dynamics of hadron production in high energy scattering processes. The FF $D_q^h(z)$, describing the probability for a parton $q$ to emit a hadron $h$ with a certain fraction $z$ of the parent parton momentum, is a crucial input to the factorization theorem for hadron production. For example, one needs unpolarized FFs for an analysis of electron-positron single inclusive annihilation into hadrons (SIA), semi-inclusive deeply inelastic scattering (SIDIS), and hadron hadroproduction \cite{ref1,ref2,ref3,ref4,ref5,ref6,ref7,ref8,ref9,ref10,ref11}. SIA may be the cleanest process in theory for extracting FFs, since knowledge of parton distribution functions is not required for a computation of its cross section \cite{ref12}. Experimental data from SIDIS multiplicities and hadron-hadron collisions provide a way to determine the flavor decomposition into quark and antiquark FFs \cite{ref13}. Light parton FFs for light mesons at a low energy scale have been calculated in effective models recently, such as the Nambu-Jona-Lasinio (NJL) model \cite{ref14,ref15} and the nonlocal chiral quark model \cite{ref16}. Data for pion and kaon productions in SIA at the $Z$ boson mass scale have been available from TASSO \cite{ref17,ref18,ref19}, TPC \cite{ref20}, HRS \cite{ref21}, TOPAZ \cite{ref22}, SLD \cite{ref23}, ALEPH \cite{ref24}, OPAL \cite{ref25}, and DELPHI \cite{ref26,ref27}. Global fits of FFs for light hadrons have been performed by several groups: FFs were extracted from fits to measured cross sections of electron-positron annihilation in \cite{ref28}, and of electron-positron annihilation, SIDIS and proton-proton collision in \cite{ref29,ref30,ref31}.

As to heavy quark FFs for heavy hadron production, Bjorken made the first theoretical attempt using a naive quark-parton model \cite{ref32}. Suzuki proposed a simple model \cite{ref33,ref34} similar to leading-order perturbative QCD (pQCD) formalism \cite{ref35}, in which the fragmenting process is factorized into the convolution of a parton-level splitting kernel with a nonperturbative heavy hadron distribution amplitude. This approach was exteded to the next-to-leading order (NLO) in \cite{ref36}, whose results agree with the data from CLEO \cite{ref37} and Belle \cite{ref38}\footnote{Only the CLEO data are available at present, and the Belle data have been removed due to an unrecoverable error in the measurement.}, and with two phenomenological models \cite{ref39,ref40} at the charm mass scale. The heavy quark FFs have been also studied in other approaches, such as the heavy quark effective theory \cite{ref41}, the potential model \cite{ref42}, and pQCD with the input of a nonrelativistic radial wave function for a heavy quarkonium \cite{ref43}.

In this paper we will apply our previous derivation of light quark and gluon FFs in the NJL model \cite{ref14,ref15} to charm quark FFs for $D$ mesons. The gluon FFs for pions and kaons from \cite{ref15} were combined with the light quark FFs \cite{ref14} in the analysis of the $e^++e^-\rightarrow h+X$ cross section, which greatly improved the consistency between theoretical results and experimental data for pion and kaon productions. The NJL model has been extended to include heavy quarks \cite{ref44}, which describes the interplay between chiral symmetry and heavy quark dynamics. To construct the charm FFs in the NJL model, we start with the evaluation of the elementary FFs at a low model scale, which is a building block for the hadronization process. The relevant quark-meson couplings are fixed by the inputs of the charm quark and $D$ meson masses in the NJL model. The jet algorithm is then implemented to simulate the whole hadronization process, from which the charm FFs are extracted at the model scale.

It is pointed out that the above charm FFs are constructed in the infinite momentum frame, namely, through many meson emissions in the jet algorithm, while the data to be compared with were collected at finite momenta, for which only the first few emissions by a parent charm quark dominate actually. We correct this mismatch by deriving a matching equation, which takes into account the finite momentum effects in one loop QCD and in parton kinematics. We first evolve the FFs from a chosen model scale to the charm scale, at which they are usually defined, and then obtain the FFs in the finite momentum frame via the matching equation. It will be demonstrated that our results for the $D$ meson production in $e^+e^-$ annihilation based on the charm FFs with the finite momentum effects accommodate well the CLEO data \cite{ref37}.

The rest of the paper is organized as follows. In Sec.~II we extract the charm FFs from the jet algorithm in the NJL model. The matching equation, which relates the charm FFs for $D$ mesons in the infinite and finite momentum frames, is derived in Sec.~III. In Sec.~IV we obtain the charm FFs including the finite momentum effects, and compute the differential cross section for $D$ meson production. Section V contains the conclusion.

\section{Charm fragmentation functions}

\begin{figure*}
 \centering
\includegraphics[scale=0.70]{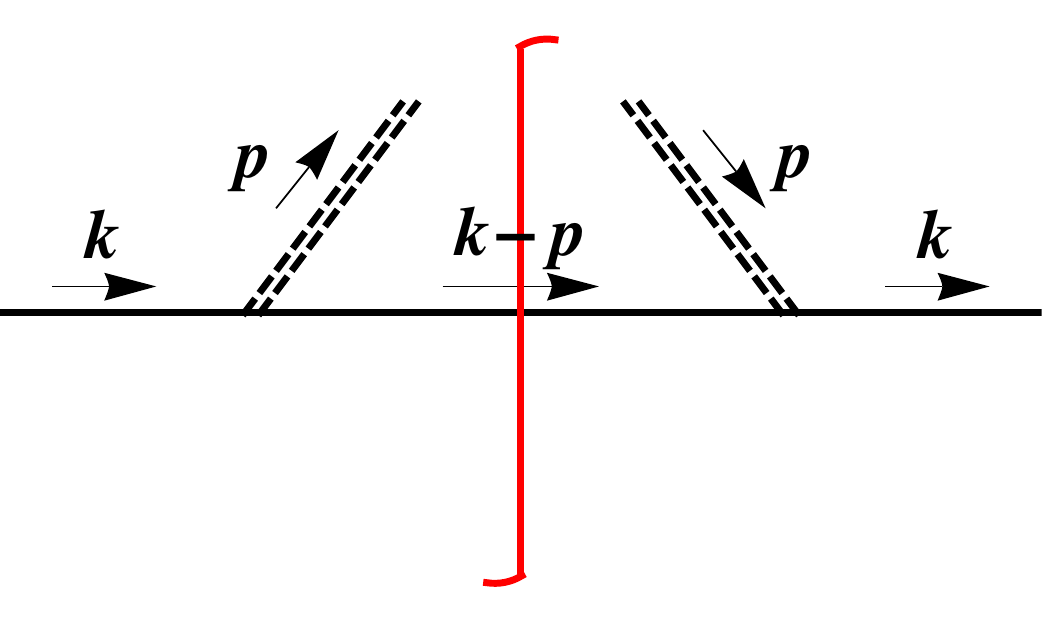}
\caption{\label{figu1}
Quark elementary FF for a pseudoscalar meson, where the solid
and dashed lines represent the quark and the pseudoscalar meson, respectively.}
\end{figure*}

We review the evaluation of the elementary FF $d_q^m(z)$ in the NJL model, which describes the probability of a single emission of the psuduscalar meson $m$ by the parent quark $q$ with the light-cone momentum fraction $z$. Its explicit expression, according to Fig.~\ref{figu1}, is written as \cite{ref13}
\begin{equation}
\begin{aligned}
d_q^{m}(z)=&-\frac{C_q^m}{2} g_{mqQ}^2 \frac{z}{2}
{\int}\frac{d^4k}{(2\pi)^4} tr\left[ S_1(k)\gamma^+
S_1(k)\gamma_5 (k\!\!\!/- p\!\!\!/+M_2) \gamma_5 \right]\\
&\times \delta(k^+-p^+/z) 2\pi \delta((k-p)^2-M_2^2)\\
=&\frac{C_q^m}{2} g_{mqQ}^2 \frac{z}{2}{\int}\frac{d^2p_\perp}{(2\pi)^3}\frac{p_\perp^2+[(z-1)M_1-M_2]^2}
{[p_\perp^2+z(z-1)M_1^2+zM_2^2+(1-z)m_m^2]^2},\\
\end{aligned}\label{eq1}
\end{equation}
where $k$ ($p$) is the parent quark (meson) momentum, $S_1$ denotes the quark propagator, $M_1$ and $M_2$ are the constituent masses of the quarks before and after the emission, respectively, and $m_m$ is the meson mass. The flavor factor $C_q^m$ depends on the composition of the meson, which takes, for example, the value 1 for $\pi^+$ and 1/2 for $\pi^0$. The dipole regulator in \cite{ref45} has been employed to avoid a divergence in the above integral. The quark-meson coupling $g_{mqQ}$ is determined via the quark-bubble graph \cite{ref13,ref45},
\begin{equation}
\begin{aligned}
\frac{1}{g_{mqQ}^2}&=-\frac{\partial \Pi(p)}{\partial p^2}\Big\vert_{p^2=m_m^2},\\
\Pi(p)=2 N_c i &\int \frac{d^4k}{(2\pi)^4} tr\left[\gamma_5 S_1(k) \gamma_5 S_1(k-p)\right],\\
\end{aligned}\label{eq2}
\end{equation}
with the number of colors $N_c$.

\begin{figure*}
 \centering
\includegraphics[scale=0.70]{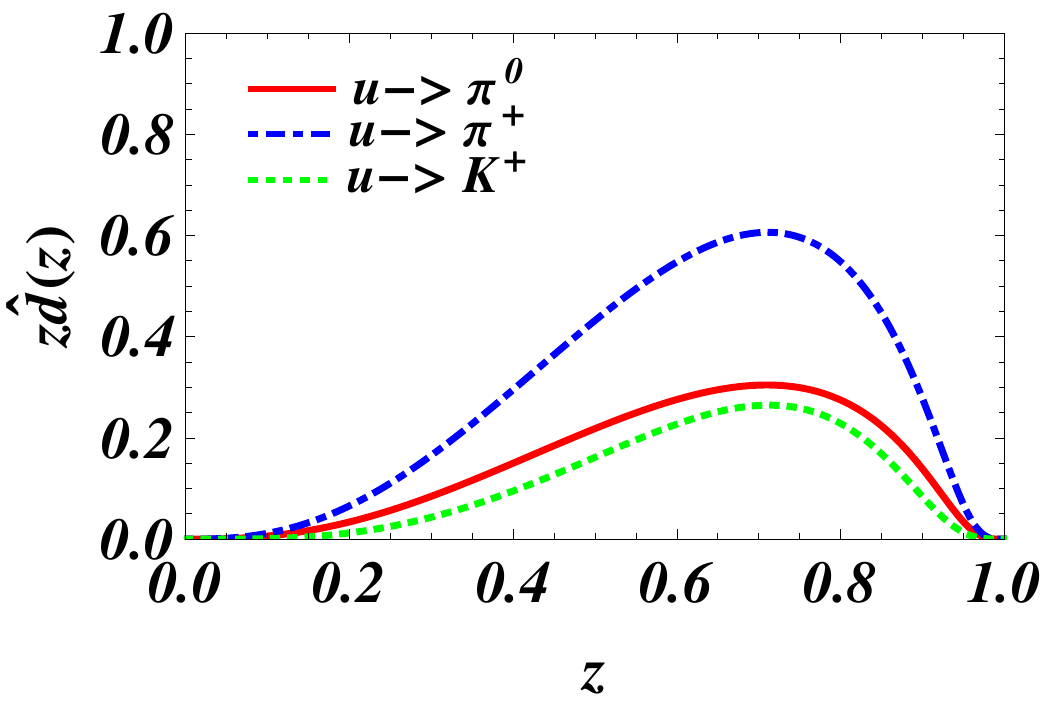}
\includegraphics[scale=0.70]{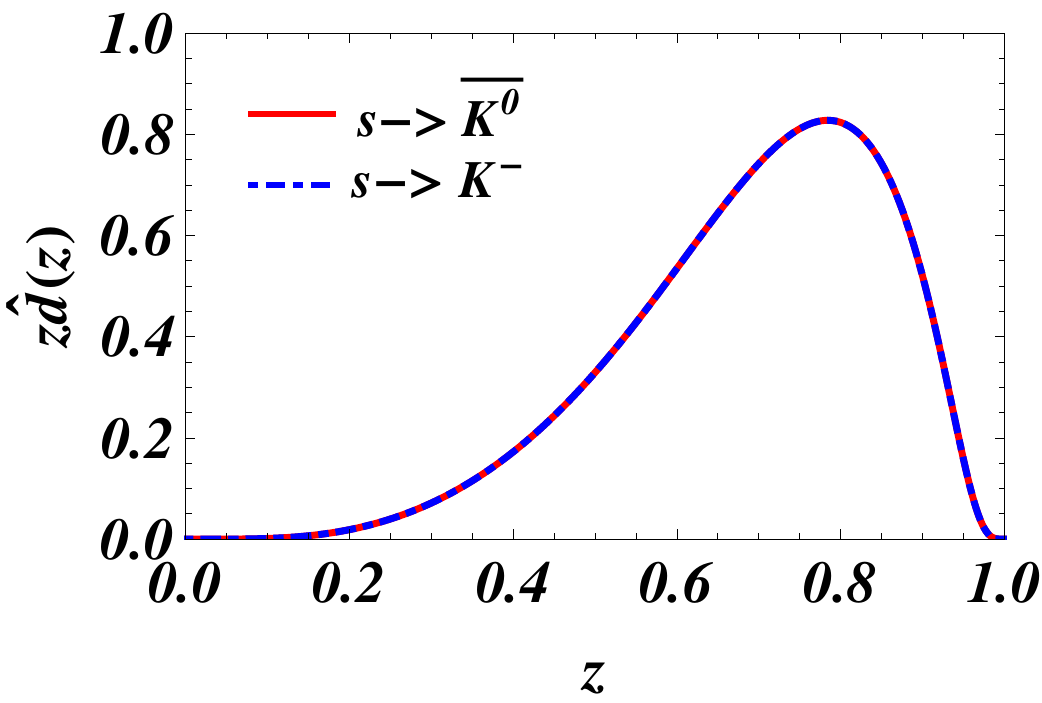}
\caption{\label{figu2}
$z$ dependencies of the $u$ quark (left) and $s$ quark (right) elementary FFs in the NJL model.}
\end{figure*}

\begin{figure*}
 \centering
\includegraphics[scale=0.70]{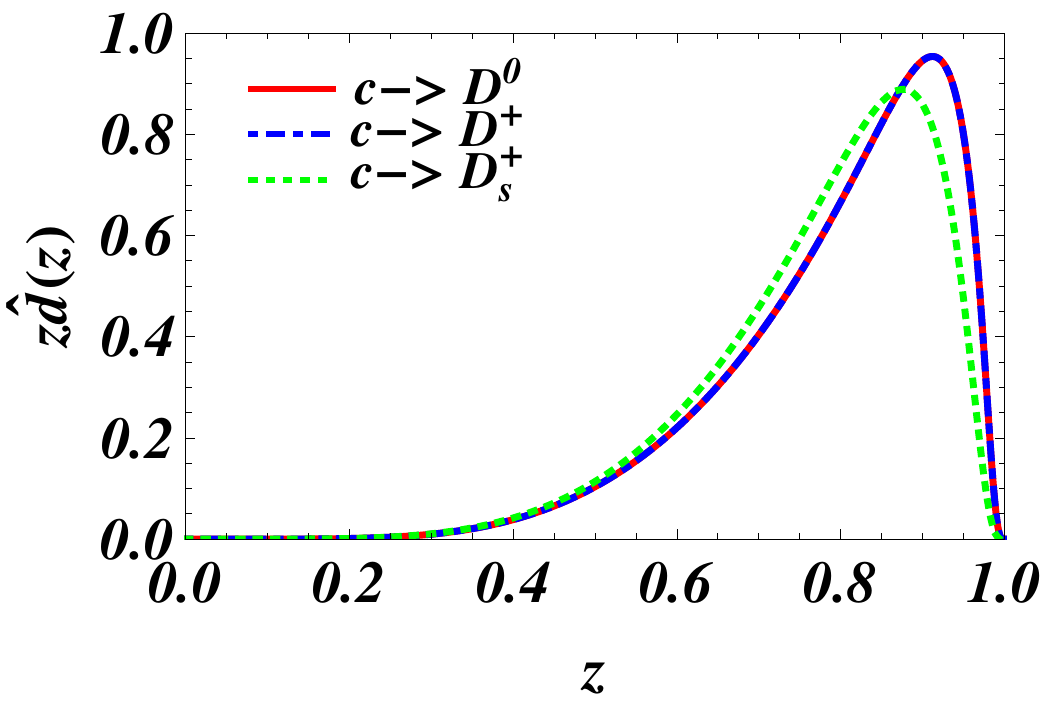}
\includegraphics[scale=0.70]{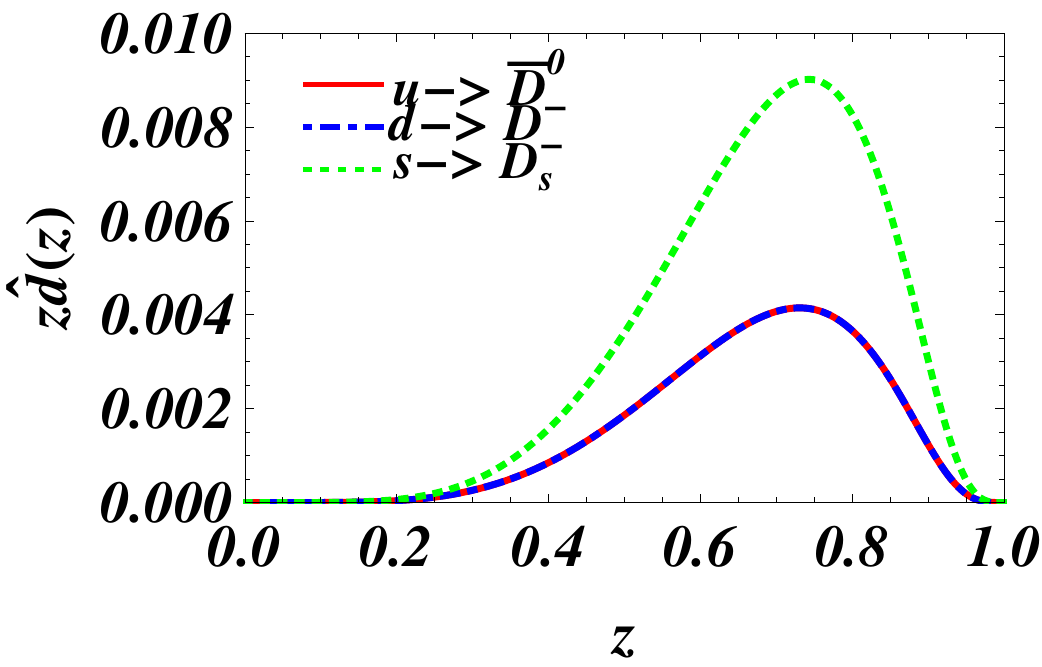}
\caption{\label{figu3}
Charm quark (left) and light quark (right) elementary FFs for $D$ mesons.}
\end{figure*}

We extend the above formalism to include charm quarks. For the parameters associated with the light quarks, we adopt $M_u=M_d=0.4$ GeV and $M_s=0.59$ GeV for the constituent quark masses, $m_\pi=0.14$ GeV and $m_K=0.495$ GeV for the meson masses, and $g_{\pi qQ}=4.24$ and $g_{KqQ}=4.52$ for the couplings fixed in \cite{ref15}. For the couplings between charm quarks and $D$ mesons, we obtain $g_{Dcu}=g_{Dcd}=1.22$ and $g_{Dcs}=1.41$ from Eq.~(\ref{eq2}) with the proper-time regularization \cite{ref46}, taking the charm quark mass $M_c=1.3$ GeV and the $D$ ($D_s$) meson mass $m_D=1.86$ ($m_{D_s}=1.96$) GeV as the inputs. We do not consider the charm fragmentation into $\eta_c$ mesons, because the corresponding coupling $g_{\eta_c cc}=0.045$
from the $\eta_c$ meson mass $m_{\eta_c}=2.98$ GeV is negligible. The $z$ dependencies of the various light quark elementary FFs and of the elementary FFs for $D$ mseons are displayed in Fig.~\ref{figu2} and \ref{figu3}, respectivly. As the initial parton is a charm quark, it must split into a $D$ meson and a light quark first. The above elementary FFs have been normalized according to the probability condition $\sum_m \int_0^1 \hat{d}_q^m(z)dz=1$ for each parent quark $q$, where the summation runs over all possible mesons $m$, including $D$ mesons. The behavior of the light quark elementary FFs in Fig.~\ref{figu2} is very close to what was obtained in \cite{ref15}, implying that the probability for light quarks to emit $D$ mesons is much lower than to emit pions and kaons, as indicated in the right plot of Fig.~\ref{figu3}. It is seen that the elementary FFs for the $c\to D^0$ and $c\to D^+$ splittings are identical, because the masses and the couplings associated with the $u$ and $d$ quarks have been set to the same values. The probability of the $c\to D_s^+$ splitting is similar to that of $c\to D^0,D^+$ due to the close quark-meson couplings and charmed meson masses. Figure~\ref{figu3} shows that a $D$ meson tends to carry a large fraction $z$ of a parent parton momentum.

\begin{figure*}
 \centering
\includegraphics[scale=0.70]{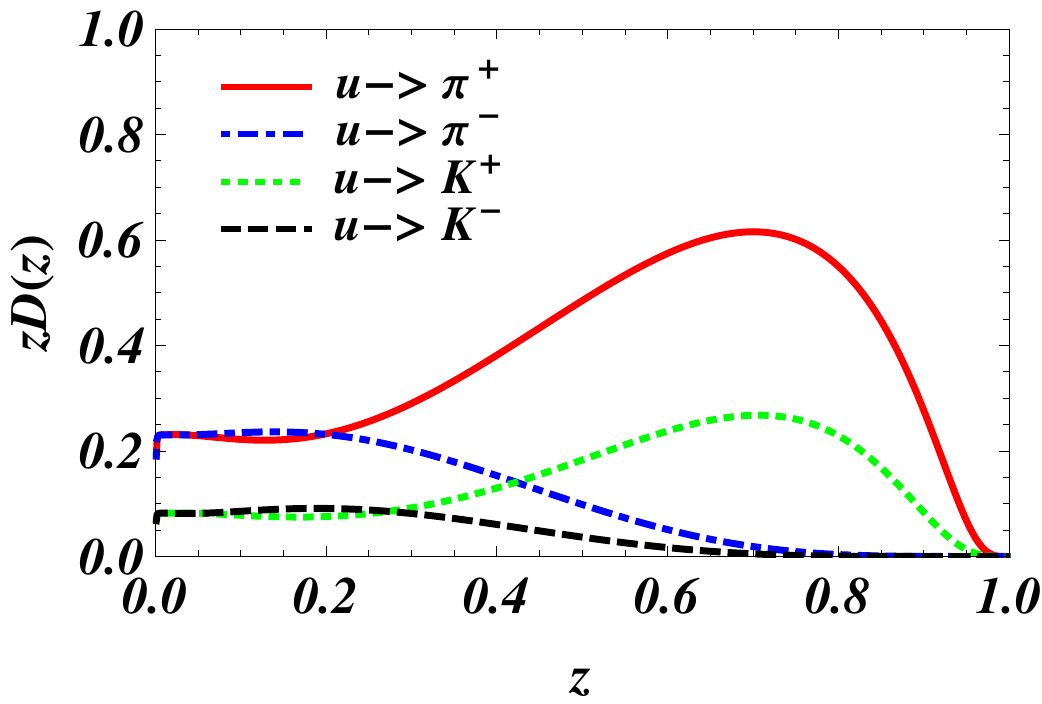}
\includegraphics[scale=0.70]{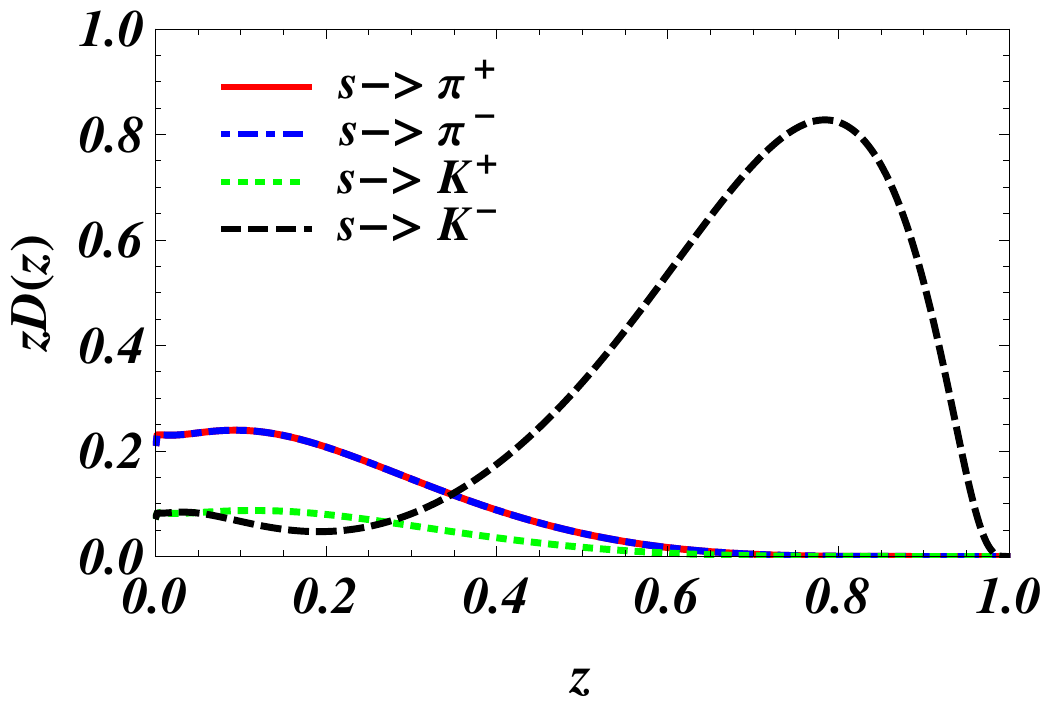}
\caption{\label{figu4}
$u$ quark (left) and $s$ quark (right) FFs for light mesons at the model scale.}
\end{figure*}

\begin{figure*}
 \centering
\includegraphics[scale=0.70]{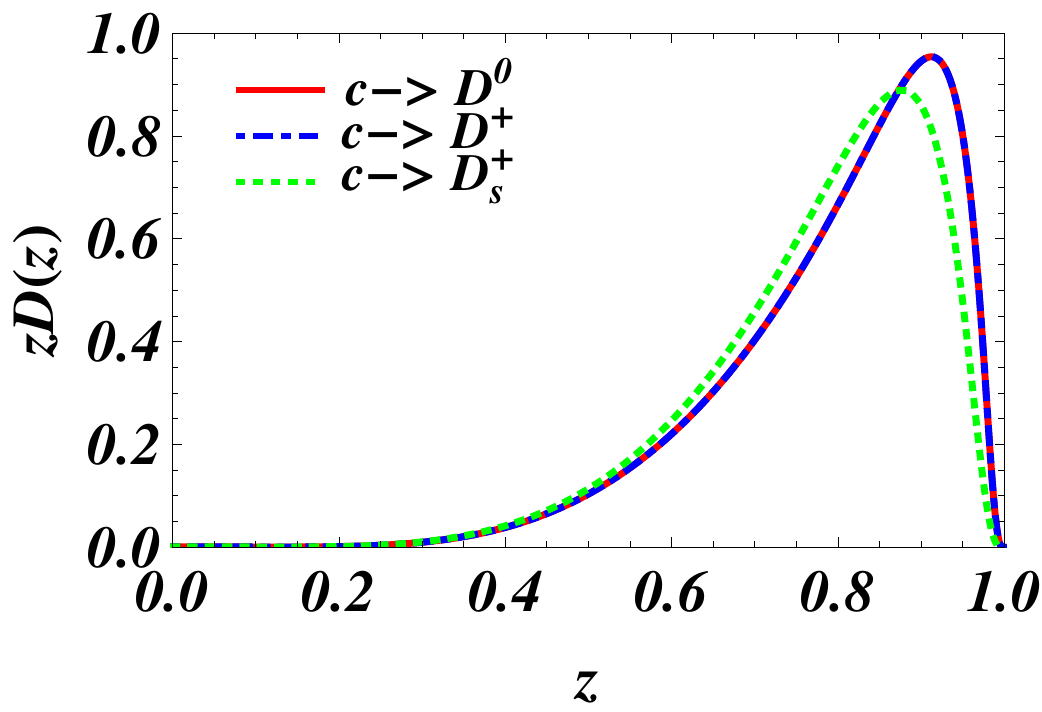}
\includegraphics[scale=0.70]{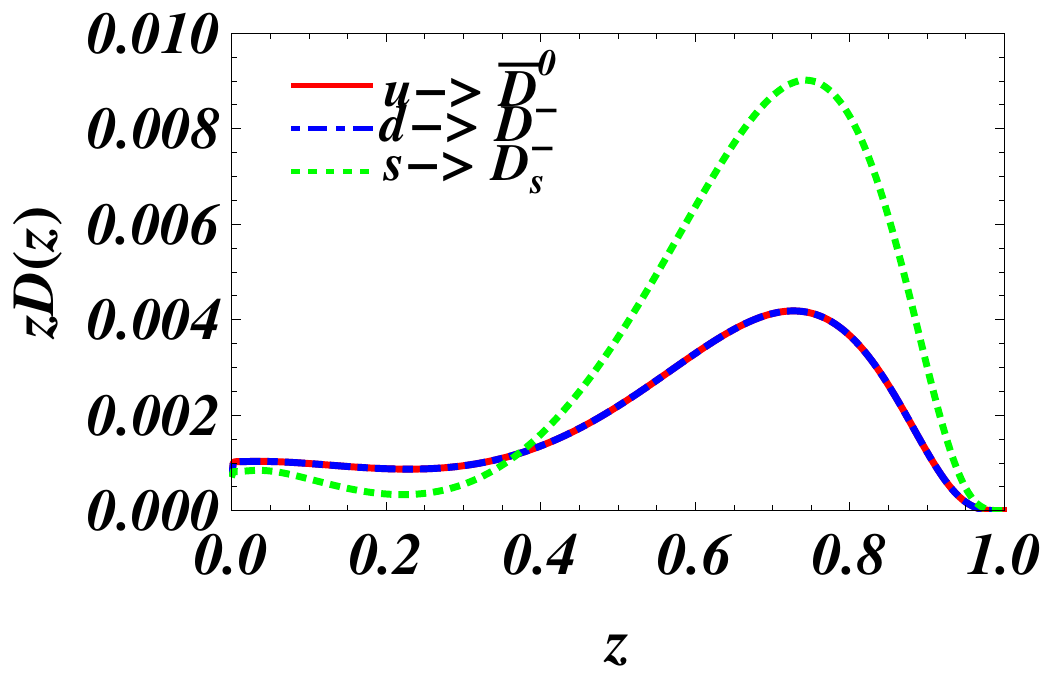}
\includegraphics[scale=0.70]{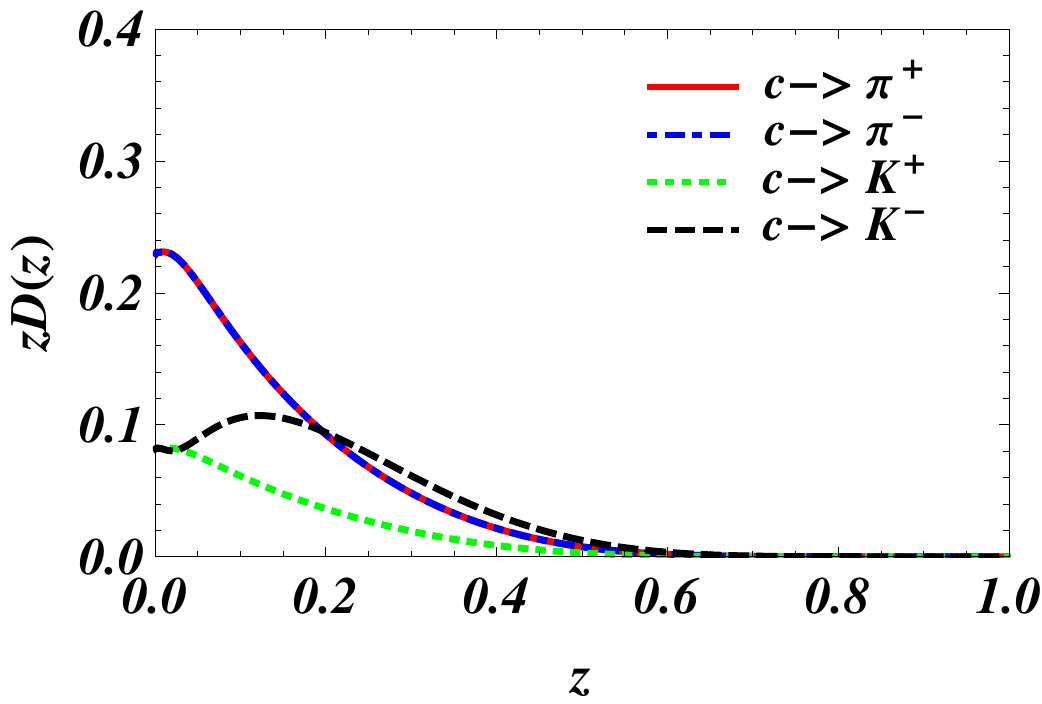}
\caption{\label{figu5}
Charm FFs for $D$ mesons (upper left), light quark FFs for $D$ mesons (upper right), and charm FFs for light mesons (lower) at the model scale.}
\end{figure*}

The integral equation based on a multiplicative ansatz for a FF is given by \cite{ref47,ref48}
\begin{equation}
\begin{aligned}
D_q^m(z)=&\hat{d}_q^m(z)+\sum_{Q}{\int_z^1\frac{dy}{y}\hat{d}_q^Q(y)D_Q^m(\frac{z}{y})},\\
&\hat{d}_q^Q(y)=\hat{d}_q^m(1-y)|_{m=q\bar{Q}}.
\end{aligned}\label{eq3}
\end{equation}
The above equation, iterating the elementary FFs to all orders, determines the probability for the quark $q$ to emit the meson $m$ with the momentum fraction $z$ through the jet process at the model scale. The first term $\hat{d}_q^m$ on the right-hand side of Eq.~(\ref{eq3}) corresponds to the first emission of the meson $m=q\bar Q$, and the second term, containing a convolution, collects the contribution from the rest of meson emissions described by $D_Q^m$ with the probability $\hat{d}_q^Q$. The extracted light quark FFs for light mesons are exhibited in Fig.~\ref{figu4}, which differ only slightly from those in \cite{ref15}, since it is difficult for light quarks to emit $D$ mesons as stated before. The charm FFs for light mesons and $D$ mesons, and the light quark FFs for $D$ mesons are presented in Fig.~\ref{figu5}. The upper left (right) plot in Fig.~\ref{figu5}, very similar to the left (right) plot in Fig.~\ref{figu3}, indicates that $D$ mesons are mainly produced at the first emission of the jet process. This explains why a $D$ meson detected in low energy experiments always carries a large momentum fraction. The upper right plot of Fig.~\ref{figu5} confirms the small probability for light quarks to emit $D$ mesons. The lower plot in Fig.~\ref{figu5} shows that light mesons carry only a small portion of a parent charm quark momentum.

As to the gluon FFs, we follow the approach in \cite{ref15}, where
a gluon is treated as a pair of quark and anti-quark in the NJL model. The gluon elementary FFs $d_g^m(z)$ are then inferred from the quark and anti-quark elementary FFs for emitting the mesons $m$ under the requirement that the quark-anti-quark pair remains in the flavor singlet state after meson emissions. We do not regard a gluon as a pair of heavy charm quarks in this work, so the gluon FFs for $D$ mesons are completely generated by QCD evolution.
It is then expected that the gluon FFs $D_g^D(z)$ for all flavors of $D$ mesons are much smaller than for light mesons, especially for pions.

\section{The QCD matching equation}

After extracting the charm quark FFs at the model scale in the previous section, we take the following steps to prepare the FFs suitable for studies of $D$ meson production in intermediate energy processes. First, we evolve the charm FFs at the model scale to the charm scale $M_c^2$, at which they are usually defined. The code QCDNUM \cite{ref49} for the NLO QCD evolution of FFs will be employed for this task. It will be observed in the next section that the evolution effect enhances the small $z$ behavior of the charm FFs, and they become nonvanishing even at $z$ as low as 0.05. The FFs from the jet process are actually constructed in the infinite momentum frame, where a parent charm quark carries a momentum much lager than $M_c$, so that many meson emissions are allowed. In reality, a charm quark produced at experiments like CLEO possesses a finite momentum, about 5 GeV at most. Hence, it is unlikely to find a $D$ meson of the mass about 2 GeV with the momentum fraction $z$ below 0.2. It hints that one needs to obtain the charm FFs defined in the finite momentum frame for practical applications.

\begin{figure*}
 \centering
\includegraphics[scale=0.4]{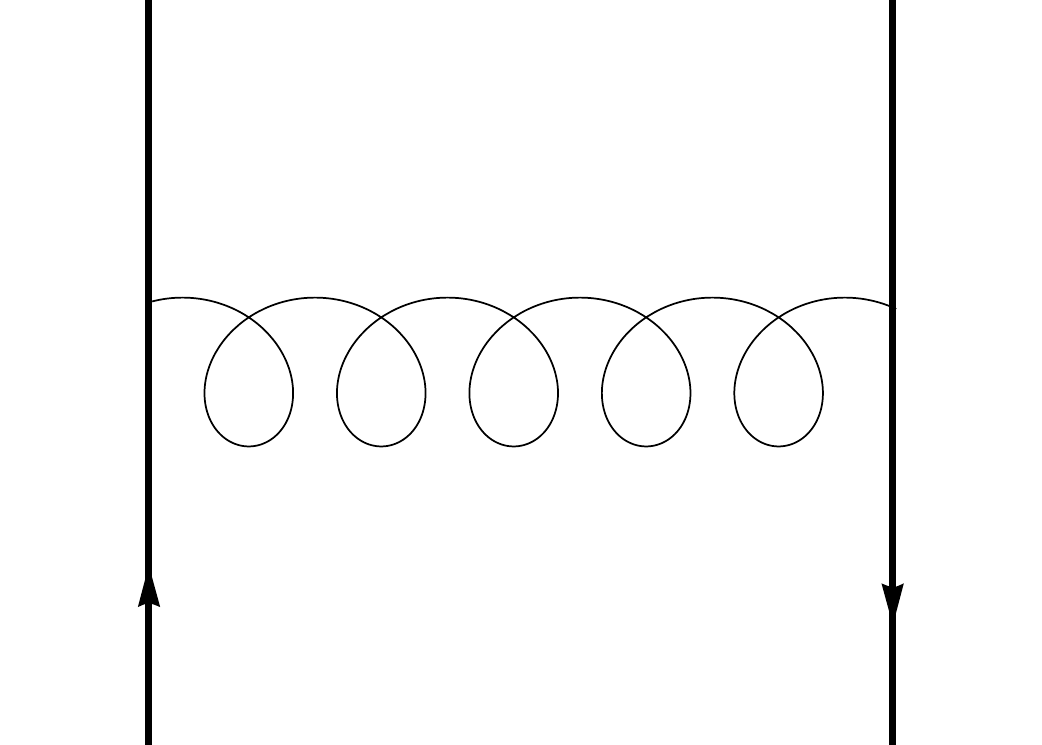}
\includegraphics[scale=0.4]{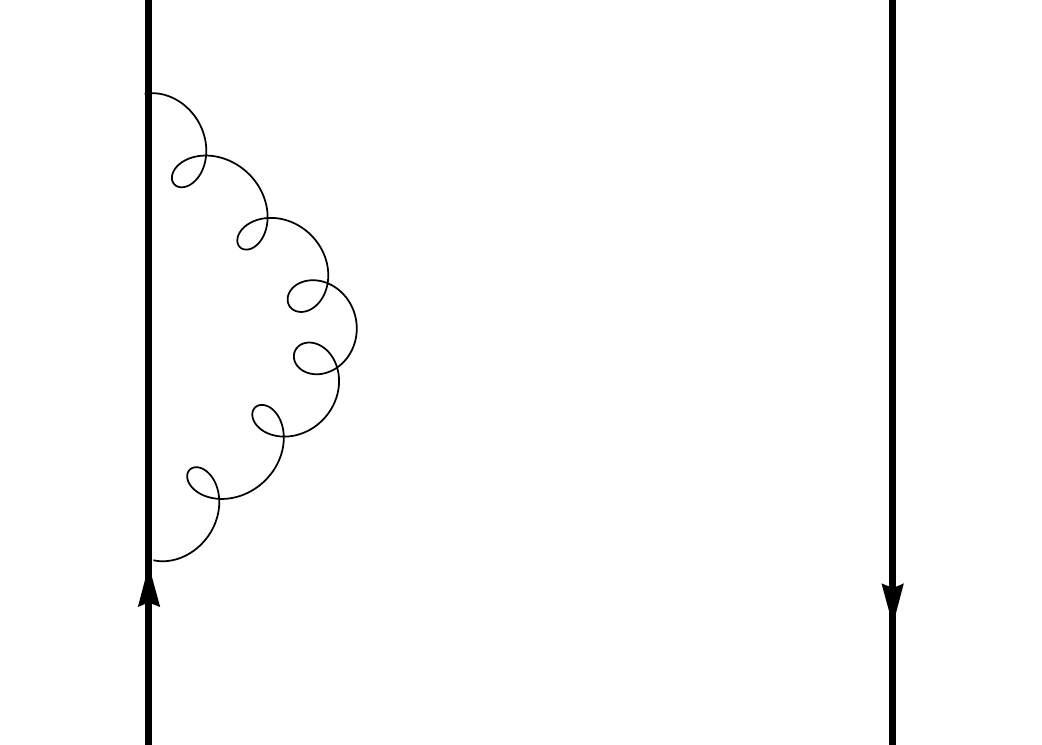}
\caption{\label{figu6}
One loop ladder diagram (left) and self-energy diagram (right) for the matching equation.}
\end{figure*}

To take into account the finite momentum effects, we derive a matching equation at one loop accuracy, which relates the charm FF $D^{\rm fin}$ in the finite momentum frame to $D^{\rm inf}$ in the infinite momentum frame,
\begin{eqnarray}
D^{\rm fin}(z)&=&\int_z^1 \frac{d \xi}{\xi} K(z/\xi)D^{\rm inf}(\xi).
\end{eqnarray}
At leading order, a parton of the momentum $p^+/z$ turns into a parton of the momentum of $p^+$ through a tree diagram. The corresponding matching kernel is written as $K^{(0)}(z)=\delta(1/z-1)$ from the momentum conservation. The calculation of the one loop matching kernel $K^{(1)}(z)$ involves the quark diagrams in Fig.~\ref{figu6}, where the ladder diagram contains a real gluon exchanged between the charm quarks before and after the final state cut. We compute these diagrams in the two frames following \cite{ref50}, and then take their difference to get $K^{(1)}(z)$. The other diagrams with gluons attaching to the Wilson lines involved in the FF definition, which give the same results in both frames, do not contribute to the matching kernel.

The loop integral for the ladder diagram in the infinite momentum frame is written as
\begin{eqnarray}
D_{\rm ladder}^{\rm inf}(z)&=&- \frac{g^2}{4}  C_F \int \frac{d^4 l}{(2\pi)^3}  \frac{\Tr [(p\!\!\!/+l\!\!\!/)\gamma^\nu
p\!\!\!/ \gamma^\nu (p\!\!\!/+l\!\!\!/) \gamma^+]}{[(p+l)^2-M_c^2]^2}
\delta(l^2)\delta(l^+-(\frac{1}{z}-1)p^+), \label{eq4}
\end{eqnarray}
where $C_F=4/3$ is a color factor, $p$ ($l$) is the momentum of the outgoing charm quark (real gluon), $z$ is the fraction relative to the incoming charm quark momentum, and the charm quark mass $M_c^2$ serves as an infrared regulator.
A straightforward evaluation yields
\begin{equation}
D_{\rm ladder}^{\rm inf}(z)
=\frac{\alpha_s}{2\pi} C_F \frac{1-z}{z}\ln \left[\frac{z^2\mu^2}{(1-z)^2 M_c^2}\right],
\label{eq5}
\end{equation}
with the ultraviolet cutoff $\mu$ for the integration over the transverse momentum $l_T$.  The ladder diagram contributes in the finite momentum frame
\begin{eqnarray}
D_{\rm ladder}^{\rm fin}(z)&=&- \frac{g^2}{4}  C_F \int \frac{d^4 l}{(2\pi)^3}  \frac{\Tr[(p\!\!\!/+l\!\!\!/+M_c)\gamma^\nu
(p\!\!\!/+M_c) \gamma^\nu (p\!\!\!/+l\!\!\!/+M_c) \gamma^+]}{[(p+l)^2-M_c^2]^2} \nonumber \\
& &\times\delta(l^2)\delta(l^+-(\frac{1}{z}-1)p^+),\nonumber\\
&=&\frac{\alpha_s}{2\pi} C_F\left\{\frac{2(1-2z)}{1-z}
+\frac{1-z}{z}\ln \left[\frac{z^2\mu^2}{(1-z)^2 M_c^2}\right]\right\},\label{eq6}
\end{eqnarray}
where all the charm mass terms have been kept. The difference between Eqs.~(\ref{eq5}) and (\ref{eq6}) defines the matching kernel from the ladder diagram
\begin{eqnarray}
K^{(1)}_{\rm ladder}(z/\xi)&=&\frac{\alpha_s}{\pi} C_F \frac{1-2z/\xi}{1-z/\xi} \nonumber\\
&=&\frac{\alpha_s}{\pi} C_F \left(\frac{1-2z/\xi}{1-z/\xi}\right)_+
+\frac{\alpha_s}{\pi} C_F \left(2- 2z+\ln\frac{\epsilon}{1-z}\right)\delta(\frac{\xi}{z}-1).
\label{eq18}
\end{eqnarray}
It is observed in the first line that the collinear divergences regularized by $M_c^2$ have cancelled between the results in the two frames. The subscript $+$ in the second line denotes a plus function, and $\epsilon$ is a soft regulator.

The self-energy diagram is calculated in the infinite momentum frame as
\begin{eqnarray}
D_{\rm self}^{\rm inf}(z)&=&-\frac{i}{4} g^2 C_F \int \frac{d^4 l}{(2\pi)^4}
 \frac{\Tr[p\!\!\!/\gamma^\nu (p\!\!\!/+l\!\!\!/)\gamma_\nu p\!\!\!/\gamma^+]}
 {(p^2-M_c^2)[(p+l)^2-M_c^2]l^2}\delta(\frac{p^+}{z}-p^+) \\
&=&-\frac{\alpha_s}{2\pi} C_F \int_0^1 dt (1-t) \ln \frac{\mu^2}{t^2M_c^2} \delta(\frac{1}{z}-1),
\label{eq16}
\end{eqnarray}
and in the finite momentum frame as
\begin{eqnarray}
D_{\rm self}^{\rm inf}(z)&=&\frac{-i}{4} g^2 C_F \int \frac{d^4 l}{(2\pi)^4}
 \frac{\Tr[(p\!\!\!/+M_c)\gamma^\nu (p\!\!\!/+l\!\!\!/+M_c)\gamma_\nu (p\!\!\!/+M_c)\gamma^+]}
 {(p^2-M_c^2)[(p+l)^2-M_c^2]l^2}\delta(\frac{p^+}{z}-p^+) \nonumber\\
&=&-\frac{\alpha_s}{2\pi} C_F \int_0^1 dt \left[1-t-\frac{2M_c^2(1+t)}{\Delta m^2}\right]
\ln \frac{\mu^2}{t^2M_c^2-\Delta m^2 t} \delta(\frac{1}{z}-1), \label{eq17}
\end{eqnarray}
where $\Delta m^2\equiv p^2-M_c^2$ will approach zero eventually.
We expand the logarithmic term in Eq.~(\ref{eq17}) in the limit $\Delta m^2\to 0$
\begin{eqnarray}
\ln \frac{\mu^2}{t^2M_c^2-\Delta m^2 t}
&=&\ln\frac{\mu^2}{t^2M_c^2}+\frac{\Delta m^2}{tM_c^2}+...\ .
\label{eq20}
\end{eqnarray}
The first term with the ultraviolet cutoff $\mu$, representing the mass correction of the charm quark, can be absorbed into the redefinition of the charm mass. The second term, removing the denominator $\Delta m^2$ in Eq.~(\ref{eq17}), produces a soft divergence.
The difference of Eqs.~(\ref{eq16}) and (\ref{eq17}) defines the self-energy contribution to the matching kernel
\begin{eqnarray}
K^{(1)}_{\rm self}(z/\xi)
&=&\frac{\alpha_s}{\pi} C_F (1-\ln\epsilon) \delta(\frac{\xi}{z}-1).
\label{eq21}
\end{eqnarray}

Combining Eqs.~(\ref{eq18}) and (\ref{eq21}), we get the one loop matching kernel
\begin{equation}
K^{(1)}(z/\xi)=\frac{\alpha_s}{\pi} C_F \left\{\left(\frac{1-2z/\xi}{1-z/\xi}\right)_+
+[3-2z-\ln(1-z)]\delta(\frac{\xi}{z}-1)\right\},
\label{eq211}
\end{equation}
where the scale of the coupling $\alpha_s$ in $K^{(1)}$ is set to $M_c$.
It is found that the soft regulator $\epsilon$ has disappeared in the sum of the ladder and self-energy diagrams, and the matching kernel is infrared finite as it should be.

\section{Differential cross section}

In addition to the matching between the QCD contributions to the charm FFs in the infinite and
finite momentum frames, the transformation between the momentum fractions in the two frames need to be implemented. Consider the tree diagram, in which the momentum $p$ ($k$) of the outgoing (incoming) charm quark is assumed to be along the $z$ axis of the finite momentum frame. The momentum fraction is defined as
\begin{eqnarray}
z\equiv \frac{p^+}{k^+}=\frac{\sqrt{(p^z)^2+M_c^2}+p^z}{\sqrt{(k^z)^2+M_c^2}+k^z},\label{z}
\end{eqnarray}
for $p^z>0$, where $k^z$ has been fixed in the plus $z$ direction. The momentum fraction in the infinite momentum frame is then given, with the charm mass being neglected, by $\xi\equiv p^z/k^z$. It is easy to find from Eq.~(\ref{z})
\begin{eqnarray}
\xi=\frac{z^2(\sqrt{1+r_c^2}+1)^2-r_c^2}{2z(\sqrt{1+r_c^2}+1)}\equiv X(z),\label{xi}
\end{eqnarray}
with the ratio $r_c=M_c/k^z$. Note that $z$ is always mapped to $\xi=0$ in the infinite momentum frame for $p^z<0$. To derive the above kinematic transformation, we have expressed $\xi$ in terms of the $z$ components of the momenta, such that the physical support of $D^{\rm inf}(\xi)$ in Eq.~(\ref{xi}) remains as $0<\xi<1$. If expressing $\xi$ in terms of the zeroth components of the momenta, a nonvanishing lower bound would appear for $\xi$. 

Incorporating the kinematic transformation into the QCD matching at one loop, we arrive at the final expression of the equation
\begin{eqnarray}
D^{\rm fin}(z)&=&\int_{X(z)}^1\frac{d\xi}{\xi}
\left[\delta\left(1-\frac{\xi}{X(z)}\right)+
K^{(1)}\left(\frac{X(z)}{\xi}\right)\right]D^{\rm inf}(\xi).\label{mat2}
\end{eqnarray}
It is noticed that the matching kernel, and thus the FFs in the finite momentum frame, depend on the parent charm momentum $k^z$ through the ratio $r_c$. As stated before, $k^z$ is not much higher than the charm mass in intermediate energy experiments, such as $k^z\approx 5$ GeV at most at CLEO \cite{ref37}.
In the region with $z<r_c/(\sqrt{1+r_c^2}+1)$, we have $X(z)<0$,
which goes outside the physical support of $D^{\rm inf}(\xi)$.
Equation~(\ref{mat2}) then implies that the FF $D^{\rm fin}(z)$ stays near zero at small $z$ till $z=r_c/(\sqrt{1+r_c^2}+1)\approx 0.2$ ($z\approx 0.1$) for $k^z\approx 3$ ($k^z\approx 5$) GeV, and that the kinematic transformation squeezes the charm FF toward high $z$, making its distribution narrower.

We remind that the momentum of a charm quark produced in experiments is not a constant, but variable. In principle, one should convolute a hard kernel for charm quark production at some momentum with the charm FFs corresponding to the same momentum, as computing a cross section. However, it is too difficult to implement such a convolution in a numerical analysis. A more realistic treatment is to obtain the charm FFs averaged over the possible range of charm quark momenta for experiments, and adopt them in the convolution. For the CLEO experiment, whose data will be compared with, the reasonable range may be
1 GeV $<k^z<3.5$ GeV, because events with vanishing and maximal $D$ meson momenta
are rare. We select the values of $k^z$ with the interval 0.5 GeV in the above
range, get the corresponding charm FFs in the finite momentum frame, and take their average
with equal weights. It has been checked that other choices of the average range centering at $k^z\sim 2$-2.5 GeV yield similar results.

\begin{figure*}
 \centering
\includegraphics[scale=0.70]{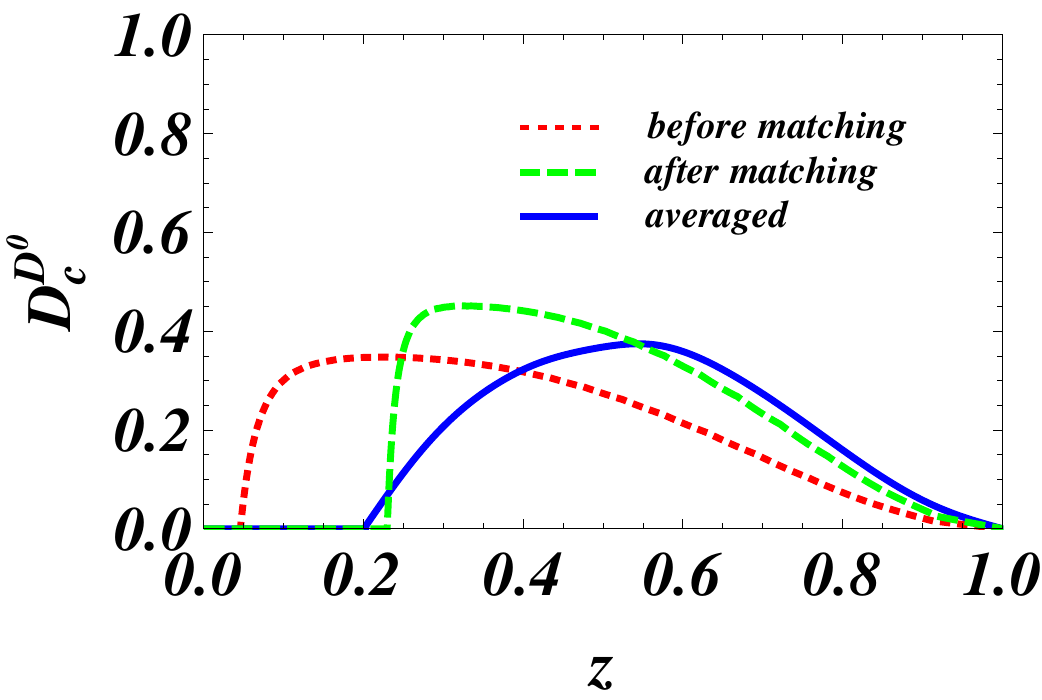}
\includegraphics[scale=0.70]{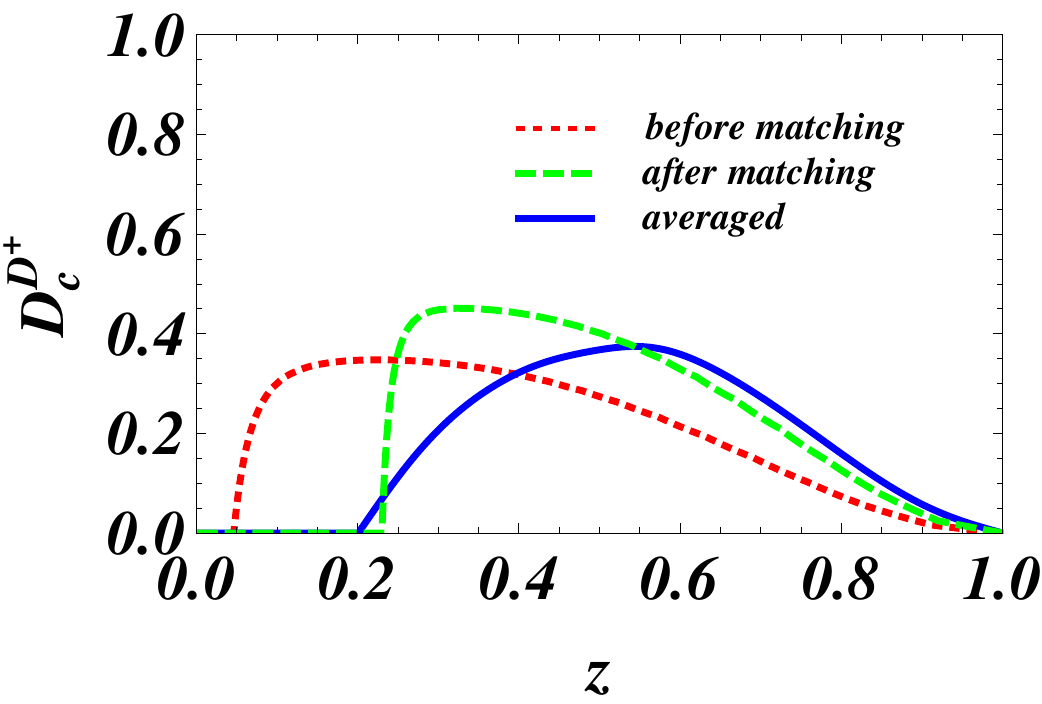}
\caption{\label{figu7}
$z$ dependencies of the charm FFs $D_c^{D^0}(z)$ (left) and $D_c^{D^+}(z)$ (right)
before and after the matching, and of the averaged charm FFs at the scale $M_c^2$.}
\end{figure*}

The model scale $Q_{0D}^2$ for the charm FFs is expected to be close to $Q_0^2$ for the light quark FFs,
but may not be exactly equal due to the inclusion of charm quarks into the NJL model.
The latter has been found to be $Q_0^2=0.17$ GeV$^2$ through the study of the pion production in $e^+e^-$
annihilation at the $Z$ boson mass scale \cite{ref15}. An ideal choice
is $Q_{0D}^2=0.15$ GeV$^2$, from which we evolve the
charm and light quark FFs for $D$ mesons to $M_c^2$ using the code QCDNUM \cite{ref49}.
The gluon FFs for $D$ mesons are generated as a consequence of the QCD evolution.
We present in Fig.~\ref{figu7} the charm FFs
in the infinite momentum frame after the NLO QCD evolution,
those converted into the finite momentum frame via the matching equation (\ref{mat2}) for $k^z=3$ GeV, and
those through the aforementioned average procedure. All the curves associated with the FFs
for $D^0$ and $D^+$ productions are almost identical as expected.
The evolution effect is quite strong, because the model scale $Q_{0D}^2=0.15$ GeV$^2$ is low:
it shifts the dominant region of the charm FFs from large $z\approx 0.9$ to small $z$.
We mention that a negative portion of the charm FFs at very small $z<0.05$, caused by the NLO evolution,
has been truncated in Fig.~\ref{figu7}. Once the charm mass is taken into
account, the light-cone component of a charm momentum does not vanish. Therefore, the combination of the
NLO matching and kinematic transformation in Eq.~(\ref{mat2}) tends to increase (decrease) the charm FFs at high (low) $z$. Specifically, it squeezes the evolved charm FFs toward the higher $z>0.2$ region for the parent charm momentum $k^z=3$ GeV.
The procedure of averaging the charm FFs in the finite momentum frame over the range 1 GeV$^2<k^z<3.5$ GeV$^2$ results in strong
suppression at small $z$, and slight enhancement at high $z$, such that the final charm FFs become more
symmetric with peaks being located at $z\approx 0.6$.

\begin{figure*}
 \centering
\includegraphics[scale=0.70]{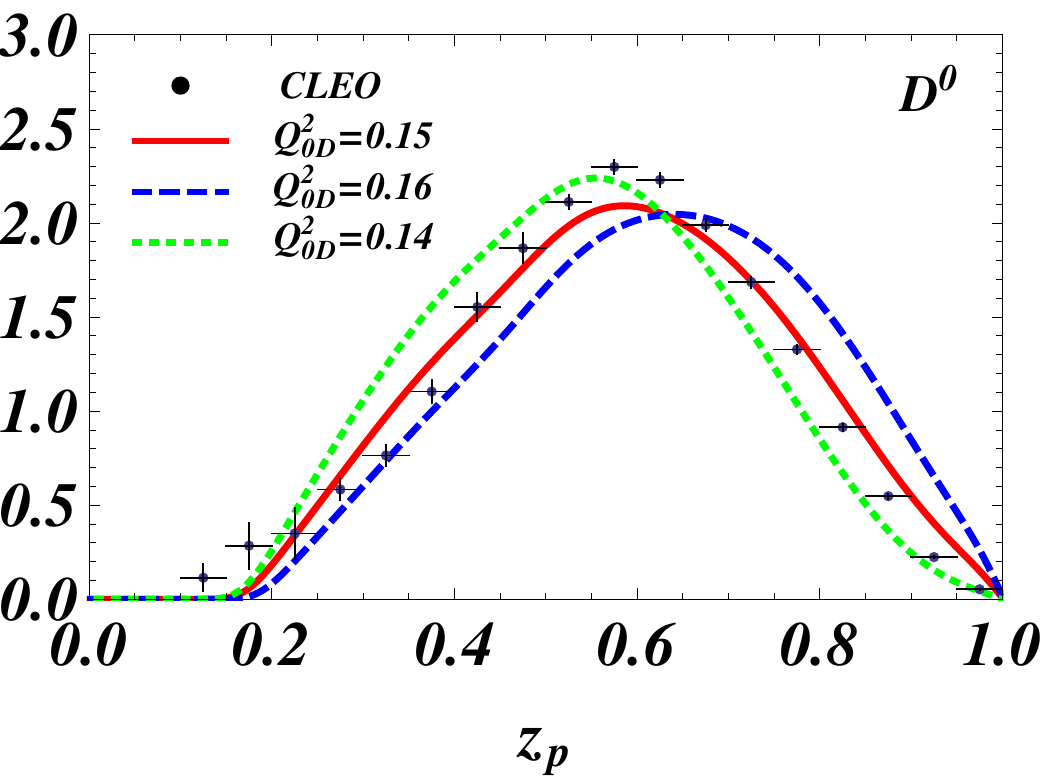}
\includegraphics[scale=0.70]{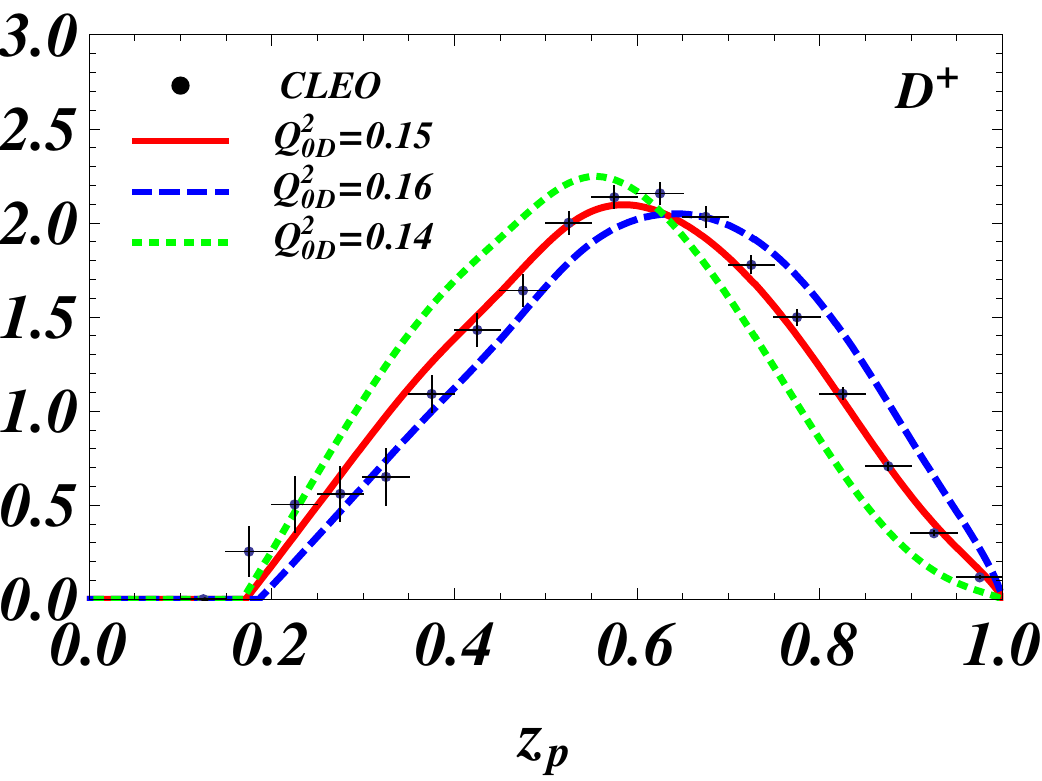}
\caption{\label{figu8}
Predicted $(1/\sigma_{\rm tot})d\sigma/dz_p$ for $D^0$ (left) and
$D^+$(right) productions with $Q_{0D}^2=0.14$, 0.15 and 0.16 GeV$^2$ at the scale $M_c^2$.
The CLEO data are also displayed for comparison.}
\end{figure*}

The differential cross section $d\sigma/dz_p$ for the $D$ meson production in $e^+e^-$
annihilation has been measured by CLEO \cite{ref37}, where the momentum fraction $z_p$ is defined by
$z_p=|{\bf p}|/|{\bf p}_{\rm max}|$, with ${\bf p}$ and ${\bf p}_{\rm max}$ being the
spatial momentum of a $D$ meson and the maximal spatial momentum observed in the measurement,
respectively. Obviously, we still need to change the momentum fraction $z$, defined in terms of the
light-cone components of $D$ meson momenta, to the variable $z_p$
in order to make comparison with the data. The former is related to the latter via
\begin{eqnarray}
z=\frac{\sqrt{(p^z)^2+m_D^2}+p^z}{\sqrt{(p^z_{\rm max})^2+m_D^2}+p^z_{\rm max}}
=\frac{\sqrt{(z_p)^2+r_D^2}+z_p}{\sqrt{1+r_D^2}+1},\label{eq8}
\end{eqnarray}
where the $D$ meson spatial momentum has been aligned with the $z$ axis, and the ratio
$r_D$ denotes $r_{D}=m_{D}/p^z_{\rm max}$ with $p^z_{\rm max}=4.95$ GeV.
We calculate the differential cross section by convoluting the hard kernel
with the averaged charm FFs in the finite momentum frame, as well as with
the light quark and gluon FFs for $D$ mesons.
For the latter, we do not employ the averaged FFs, since their contributions are negligible.

Our results for the normalized differential cross section
$(1/\sigma_{\rm tot})d\sigma/dz_p$, $\sigma_{\rm tot}$ being the total cross section,
are displayed in Fig.~\ref{figu8}, and compared with the CLEO data. It is observed
that the consistency between our results and the data, both
of which have peaks at $z\approx 0.6$, is
satisfactory, especially for the $D^+$ meson production. This consistency supports
our choice of the model scale $Q_{0D}^2=0.15$ GeV$^2$. To test the sensitivity
to the model scale $Q_{0D}^2$, we vary it by 0.01 GeV$^2$, and
show the results corresponding to $Q_{0D}^2=0.14$ and
0.16 GeV$^2$ also in Fig.~\ref{figu8}. It is easy to understand that the peak of the differential
cross section moves toward the small $z$ region as $Q_{0D}^2$ decreases, because the QCD evolution effect
gets stronger. The lower bound of $z_p$ is basically fixed by the kinematic transformation, so that
the distribution of the normalized differential cross section becomes narrower, and the peak becomes sharper.
The zones enclosed by the three theoretical curves cover all data points roughly.

\section{Conclusion}

In this paper we have derived the charm quark FFs for $D$ mesons in the
NJL model, which describe the probability for a $D$ meson to take a fraction
$z$ of a parent charm momentum. The evaluation of the corresponding
elementary FFs and the jet algorithm for producing final state mesons were
performed by including charm quarks into the NJL model.
To confront our results with the data for $D$ meson production at finite energy,
we have first evolved the FFs from their model scale to the charm scale.
The model scale $Q_{0D}^2$ for the charm FFs, from which the QCD evolution starts, is
the only uncertain parameter in the analysis. It has been found that the favored choice
$Q_{0D}^2=0.15$ GeV$^2$ is close to $Q_{0}^2=0.17$ GeV$^2$ for the light quark FFs determined
in our previous work. The evolution effect is significant enough
to shift the dominant region of the charm FFs to lower $z$. We then obtained
the charm FFs with the finite momentum effects in one loop QCD and in the definitions of the $D$
meson momentum fraction through the matching equation. It was shown that the combined QCD
matching and kinematic transformation squeezed the charm FFs toward the larger $z$ region.

To acquire more realistic charm FFs to be input into the convolution with the
hard kernel for charm quark production, we have further taken the average of
the FFs over a possible range of $D$ meson momenta in the considered experiment.
The resultant distribution is more symmetric with a peak around $z\approx 0.6$.
The contributions to the $D$ meson production from the light quark and gluon FFs,
despite of being negligible, were also included for completeness. At last, the momentum
fraction $z$ defined in terms of the light-cone components of $D$ meson momenta has to be transformed into $z_p$ defined
in terms of spatial momenta by experimentalists. It has been demonstrated, after
all the above nontrivial treatments, that the averaged charm quark FFs lead to
results for the $D$ meson production in $e^+e^-$ annihilation in agreement with
the CLEO data. We have examined the sensitivity of our results to the tunable model scale,
and observed that the variation of $Q_{0D}^2$ within 0.14-0.16 GeV$^2$
could accommodate the CLEO data well.

We emphasize the potential applications of the matching equation derived in Sec.~III.
It may not be accurate to apply the usual light-cone FFs defined in the
infinite momentum frame to analyses of intermediate energy processes,
especially when collision energy is not much higher than the mass of
a produced heavy quark. Our matching equation relates the FFs at low momenta
to those at high momenta by taking into account the finite momentum
effects in QCD and in parton kinematics. In this sense partial higher power
corrections to the factorization theorem of intermediate energy processes
have been taken into account. The FFs after the above matching should be
more suitable for studies of heavy particle production
at intermediate energy, such as that in Belle experiments. We will investigate this
subject in detail elsewhere.

\begin{acknowledgments}

This work is supported in part by
the Ministry of Science and Technology of R.O.C. under
Grant No. MOST-107-2119-M-001-035-MY3.
\end{acknowledgments}


\end{document}